\documentclass [11pt]{article}

\usepackage{amsmath}
\usepackage{amsfonts}
\usepackage{amssymb}
\usepackage{graphicx}
\usepackage{xcolor}
\usepackage{ulem}
\usepackage{graphicx}
\usepackage[body={6.5in, 9in}, right=1in, top=1in]{geometry}
\usepackage{amssymb}
\usepackage{amsmath} 
\usepackage[numbers,sort&compress]{natbib}

\makeatletter
\def\section{\@startsection {section}{1}{\z@}{-3.5ex plus -1ex minus
 -.2ex}{2.3ex plus .2ex}{\large\bf}}
\def\subsection{\@startsection{subsection}{2}{\z@}{-3.25ex plus -1ex
minus -.2ex}{1.5ex plus .2ex}{\normalsize\bf}}
\makeatother
\makeatletter

\@addtoreset{equation}{section}

\makeatother

\def\be{\begin{equation}}
\def\ee{\end{equation}}

\newcommand\eea{\end{eqnarray}}
\newcommand\bea{\begin{eqnarray}}

\newcommand{\nn}{\nonumber}

\def\({\left(}
\def\){\right)}

\usepackage{soul}
\newcommand{\Comment}[1]{{}}
\definecolor{MyDarkBlue}{rgb}{0.15,0.15,0.45}
\usepackage[linktocpage=true]{hyperref}
\hypersetup{
colorlinks=true,
citecolor=MyDarkBlue,
linkcolor=MyDarkBlue,
urlcolor=MyDarkBlue,
}

\begin{document}
\def\thefootnote{\fnsymbol{footnote}}

\hfill SISSA 46/2015/FISI \\

\begin{center}
\Large{\textbf{Inequivalence of Coset Constructions for Spacetime Symmetries:  Coupling with Gravity}} \\[0.5cm]
 
\large{Pietro Baratella$^{\rm a,b}$,  Paolo Creminelli$^{\rm c}$, Marco Serone$^{\rm a,b,c}$ and Gabriele Trevisan$^{\rm a,b}$}
\\[0.5cm]

\small{
\textit{$^{\rm a}$ SISSA, via Bonomea 265, 34136, Trieste, Italy}}

\vspace{.2cm}

\small{
\textit{$^{\rm b}$ INFN - Sezione di Trieste, 34151 Trieste, Italy}}

\vspace{.2cm}

\small{
\textit{$^{\rm c}$ Abdus Salam International Centre for Theoretical Physics\\ Strada Costiera 11, 34151, Trieste, Italy}}

\vspace{.2cm}

\end{center}

\vspace{.8cm}

\hrule \vspace{0.3cm}
\noindent \small{\textbf{Abstract}\\
We study how the coupling with gravity of theories with non-linearly realized space-time symmetries is modified when one changes the parametrization of the coset. As an example, we focus on the so-called Galileon duality: a reparametrization which maps a Galilean invariant action into another one which enjoys the same symmetry. Starting with a standard coupling with gravity, with a parametric separation between the Planck scale and the typical scale of the coset, one ends up with a theory without such a separation. In particular an infinite set of higher-dimension operators are relevant when the superluminality of the Galileon is measurable in the effective theory. This addresses an apparent paradox since superluminality arises in the dual theory even when absent in the original one.
} 
\vspace{-0.1cm}
\noindent
\hrule
\def\thefootnote{\arabic{footnote}}
\setcounter{footnote}{0}

\section{Introduction and motivation}

The generalization of the Callan, Coleman, Wess and Zumino (CCWZ)  coset construction \cite{Coleman:1969sm,Callan:1969sn}  to spacetime symmetries,  developed in refs.~\cite{Volkov:1973vd,Ivanov:1975zq}, is not as well understood as its counterpart for internal symmetries.  In particular, it is still not completely established whether different parametrizations of the coset are physically equivalent or not.

This issue has recently been addressed in ref.~\cite{Creminelli:2014zxa}, by focusing on Galileon theories \cite{Nicolis:2008in}, i.e.~on the coset Gal$(3+1,1)/$ISO$(3,1)$. In this coset one can choose different parametrizations which are isomorphic to each other: the mapping from one parametrization to another (which amounts to a field redefinition and a field-dependent change of coordinates) maps a Galilean invariant theory to another one with the same symmetry. These mappings were dubbed Galileon duality \cite{deRham:2013hsa,Kampf:2014rka}. A relevant feature of the duality, which challenges the equivalence of different representations, is superluminality: starting with a free theory the duality gives a theory where superluminal motion with respect to the Minkowski light cone is possible and measurable within the regime of validity of the effective theory. The same phenomenon was already observed in the case of non-linear realizations of the conformal group \cite{Creminelli:2013fxa}.  

In ref.~\cite{Creminelli:2014zxa} it was shown that different parametrizations are non-locally related to each other and in this sense are inequivalent.  Indeed the coupling to external sources in the two theories is necessarily different: if local in one parametrization, it has to be non-local in the other one. This non-locality comes to a rescue when superluminality is addressed: when superluminality with respect to the Minkowski light cone is observed, the coupling with sources is effectively non-local so that no paradox is generated. This interpretation is however not conclusive, since it does not address what happens when other dynamical fields are added: in particular, if these fields transform under the coset reparametrization,
the mapped theory could still be local in the new coordinates.
From this point of view all fields after the duality might be sensitive to an effective metric induced by the reparametrization, so that the question of superluminality with respect to the Minkowski light cone in the new theory would become immaterial. 

Aim of this note is to extend the work of ref.~\cite{Creminelli:2014zxa} by reconsidering the mapping in the presence of other dynamical fields. In particular we couple gravity to the coset construction, using the recent formulation of ref.~\cite{Delacretaz:2014oxa}. The main reason to do so is that gravitons are expected to move on the light cone, at least if the coupling of gravity is sufficiently standard, so that the issue of superluminality can be formulated more physically in terms of motion with respect to the graviton. Let us stress that we are not studying here the Galileon as a limit of massive gravity (or bigravity) \cite{deRham:2014zqa,deRham:2014lqa}.
In that case  the galileon duality transformation is essentially a diffeomorphism and the coupling of $\pi$ with gravity is non-standard, since $\pi$ describes the (longitudinal) fluctuations of the space-time dependent Stueckelberg fields \cite{deRham:2014zqa}.  In our case, we study ordinary Einstein gravity with minimally coupled $\pi$ and the reparametrization is not a diffemorphism. For us the Galileon is just an example of a space-time coset: we expect the same arguments to be straightforwardly extended to other cosets, for example to the non-linear realization of the conformal group.

In section \ref{sec:cosets} we briefly review the mapping found in ref.~\cite{Creminelli:2014zxa} and discuss the transformation properties of additional fields. In section \ref{sec:gravity} we generalize the mapping in the presence of gravity: we show that ordinary Einstein-Hilbert gravity in one coset parametrization is mapped by the duality to a complicated theory where torsion is induced (section \ref{sec:torsion}).
We explicitly write the action in the other parametrization and show that the induced torsion pollutes the action with an infinite number of higher-derivative terms (section \ref{sec:dualaction}). All these higher-derivative terms become relevant in the regime when superluminality is measurable, so that the theory is effectively non-local in this regime. 
We conclude in section \ref{sec:conclusions} stressing that different representations are inequivalent once gravity is included, since a standard gravity coupling is mapped to a non-standard one, where an infinite set of higher-dimension operators becomes relevant at scales much lower than the Planck scale.

\section{Galileon Cosets}
\label{sec:cosets}

The algebra of the Galileon group Gal(3+1,1) consists of the Poincar\'e algebra plus the generators $B_a$ and $C$. Their commutation relations are given by \cite{Goon:2012dy}
\be\begin{split}
\left[ M_{a b},B_{c} \right] & =\eta_{ac}B_{b}-\eta_{bc}B_{a} \,, \\
\left[ P_a,B_b\right] & =\eta_{ab}C \,,
\end{split}
\end{equation}
where $P_a$ and $M_{ab}$ are the translation and Lorentz generators, respectively.
The other commutation rules are either well known or trivial.

We are interested in theories where the Galileon group is spontaneously broken down to the Poincar\'e group. The low-energy description of the Goldstone modes is captured by the coset construction \cite{Coleman:1969sm,Callan:1969sn}, suitably generalized for spacetime symmetries \cite{Volkov:1973vd,Ivanov:1975zq}. The coset can be parametrized by
\be
\begin{split}
g(y^\mu,\pi,\Omega^\mu)& =e^{y^\mu P_\mu}e^{\pi C}e^{\Omega^\mu B_\mu} \;.
\end{split}
\ee
The Galileon algebra has a group of automorphisms, which is isomorphic to $GL(2,\mathcal{R})$ \cite{Kampf:2014rka}. We here focus on the subgroup (isomorphic to $\mathcal{R}$) given by the transformation $B \rightarrow B+\alpha P\equiv B_{(\alpha)}$.\footnote{The other relevant generator of the group is given by $P \rightarrow P+\beta B \equiv P_{(\beta )}$ (the other transformations just correspond to rescaling of the coordinates and fields). This transformation, parametrized by $\beta$, corresponds to studying the perturbations of the original theory around a background of the form $\pi = \beta x^2/2$: the action for perturbations is still Galilean invariant \cite{Nicolis:2008in}. In this case gravity breaks the equivalence between the two parametrizations in a rather trivial way, since one has to choose whether it is $P$ or $P_{(\beta)}$ which defines the coupling with gravity. In more physical terms, the background one is expanding around gravitates. Notice that in the transformation we consider here $P$ is left untouched.}
The coset is now represented as
\begin{equation}
\begin{split}
g_{\alpha}(x^\mu,q,\hat{\Omega}^\mu) & =e^{x^\mu P_\mu}e^{q C}e^{\hat{\Omega}^\mu B_{(\alpha)\mu}} \,.
\end{split}
\end{equation}
After imposing the inverse Higgs constraint \cite{Ivanov:1975zq}, the Cartan forms in the two parametrizations read 
\begin{equation}
\begin{split}
g^{-1} \text{d}g= & \;  \text{d}y^\mu P_\mu+\text{d}\Omega^\mu B_\mu \,, \\
g^{-1}_\alpha \text{d}g_\alpha= & \, \Big( \text{d}x^\mu+\alpha\text{d}\hat\Omega^\mu \Big) P_\mu+ \text{d}\hat \Omega^\mu B_\mu \,,
\label{Cartan}
\end{split}
\end{equation}
where
\be
\Omega_\mu(y)=-\frac{\partial \pi}{\partial y^\mu}  \,, \ \ \ \ \ \ \hat \Omega_\mu(x)=-\frac{\partial q}{\partial x^\mu}  \ .
\ee
By equating the two Cartan forms,  the following relations are found \cite{Creminelli:2014zxa}:\footnote{The representatives chosen here differ by a trivial rescaling from that taken in ref.~\cite{Creminelli:2014zxa}. The parameter $\alpha$, with dimensions $[{\rm energy}]^{-3}$, equals $\alpha=1/\Lambda^3$.}  
\bea	\label{eq.coordtransf}
y^\mu &= &x^\mu-\alpha\partial^\mu q(x)\ ,\\
\label{eq.dqdpi}
\frac{\partial \pi(y)}{\partial y^\mu} &= &\frac{\partial q(x)}{\partial x^\mu} \ ,\\
\pi(y)&= & q(x)-\frac{1}{2}\alpha (\partial q(x))^2 \,.
\label{eq.piq}
\eea
The Goldstone bosons $\pi$ and $q$ transform in the same way under the non-linearly realized $B_\mu$ and $C$ transformations, in their respective coordinates:
\be
\delta \pi(y) = c + b_\mu y^\mu\,, \ \ \ \ \ \ \delta q(x) = c + b_\mu x^\mu\,.
\ee
How do other fields, inert under the Galilean symmetry, transform under this change of representation? 
Since we are redefining the coordinates, such fields will transform as under a diffeomorphism \cite{deRham:2014lqa}.
For example, a scalar $\psi$ transforms as
\be\label{eq.psitrans}
\hat\psi(x) \equiv \psi (y(x)) \;.
\ee 
Let us show in simple examples how the transformation (\ref{eq.psitrans}) ensures  that the resulting action is invariant under the Galilean symmetry.
Consider for instance the Galilean invariant interaction
\be
\int {\rm d}^4 y \;\psi(y) \Box \pi(y) \;.
\ee
Using eqs.~\eqref{eq.coordtransf}, \eqref{eq.dqdpi} and \eqref{eq.psitrans} we have
\be
\int {\rm d}^4 x \left|\det\left(\frac{\partial y}{\partial x}\right)\right| \hat\psi (x) \frac{\partial x^\nu}{\partial y^\mu} \frac{\partial}{\partial x^\nu} \frac{\partial q}{\partial x^\mu} \;.
\ee
The product of the determinant of the Jacobian and the matrix $\partial x/\partial y$ gives a (finite) number of objects of the form $\partial\partial q$, so that the resulting action is of the schematic form
\be\label{eq.tildepsiact}
\int {\rm d}^4 x \;\hat\psi(x) (\partial\partial q+ \alpha (\partial\partial q)^2 + \alpha^2 (\partial\partial q)^3 + \ldots ) \,,
\ee
which is obviously Galilean invariant. If we had not performed the $\psi$ field redefinition, we would have had 
\be
\int {\rm d}^4 x \;\psi(x -\alpha \partial q) (\partial\partial q+ \alpha (\partial\partial q)^2 + \alpha^2 (\partial\partial q)^3 + \ldots ) \;.
\label{NotInv}
\ee
The action (\ref{NotInv}) is {\it not} Galilean invariant: for instance the only new term compared with the action \eqref{eq.tildepsiact} with four derivatives is $\partial\psi\partial q \Box q$, which is not invariant under a shift of $\partial q$. One can check that the same argument works for fields which are not scalars: in order to preserve Galilean invariance in the new frame, they must transform in the same way as under a diffeomorphism. 

At this point everything looks almost trivial: all fields are transformed from the $y$ to the $x$ coordinates and in particular the Minkowski light cone in the original $\pi$ picture is mapped to a new $q$-dependent surface. From this point of view it seems there is nothing strange in having motions which are on the light cone in the $\pi$ picture and become sub- or superluminal after the duality. However this is too quick. We said that extra fields must transform as under a diffeomorphism to generate a new Galilean invariant action. Exactly for the same reason the metric (let us take it non-dynamical for the time being) {\it must not} transform as under a diffeomorphism. The metric $\eta_{\mu\nu}$ does not change under the duality: both in the $y^\mu$ and $x^\mu$ coordinates, indices are contracted with $\eta_{\mu\nu}$. Let us check it in a simple example. Starting from the kinetic term $\int {\rm d}^4 y \;(\partial\pi)^2$, applying eqs.~\eqref{eq.coordtransf} and \eqref{eq.dqdpi} we get
\be
\int {\rm d}^4 x \left|\det\left(\frac{\partial y}{\partial x}\right)\right| (\partial q)^2 \;.
\ee
Writing the determinant in terms of $q$ we get a Lagrangian of the schematic form $\big(\alpha\partial\partial q+ \alpha^2(\partial\partial q)^2 + \alpha^3(\partial\partial q)^3\big) (\partial q)^2$ and one can check this is a linear combination of four Galileon interactions \cite{Nicolis:2008in}. The metric remained $\eta_{\mu\nu}$. One may be tempted to write the action in terms of the metric 
\be
g_{\mu\nu} \equiv \frac{\partial y^\rho}{\partial x^\mu} \frac{\partial y^\sigma}{\partial x^\nu} \eta_{\rho\sigma} \,,
\ee
but one should resist this temptation. Indeed in this case one would perform a standard diffeomorphism except for the peculiar transformation of the derivatives of the Galileon, eq.~\eqref{eq.dqdpi}. The resulting action would be
\be
\int {\rm d}^4 x \sqrt{-g} g^{\mu\nu} \frac{\partial y^\rho}{\partial x^\mu} \frac{\partial y^\sigma}{\partial x^\nu}  \frac{\partial q}{\partial x^\rho} \frac{\partial q}{\partial x^\sigma} \;.
\ee
This expression has two problems: i) if one expands the two $\partial y/\partial x$ in terms of $q$ the resulting operators are not Galilean invariant
(for instance one can check that there is a quartic operator which induces equations of motion with single derivatives on some of the $q$'s), ii) the second derivatives acting on $q$ are not covariant with respect to the metric $g$. We conclude that the non-dynamical metric $\eta_{\mu\nu}$ must remain invariant under the duality transformation if we want to end up with a sensible Galilean invariant action.

There is another reason why the metric has a different status compared to other fields. One expects the graviton to move on the light cone defined by the metric and in particular to move on the Minkowski light cone on scales where curvature can be neglected. This surely happens if the system is coupled to gravity with a minimal coupling (independently of possible ambiguities related to higher derivative terms) and there is a parametric separation between the Planck scale and the typical scale of the system:  in the following we are going to call this setup a standard coupling with gravity. The parametric separation assures that in the limit $M_{\rm Pl} \to \infty$ gravitons are decoupled and thus propagate on the Minkowski light cone. With a standard coupling with gravity the Minkowski light cone defines the motion of gravitational waves: how is it then possible that the relative velocity of a given particle with respect to gravitons changes in going from the $\pi$ to the $q$ picture? The only way out is that if gravity is standard in one representation, it is not standard in the other.


\section{Coupling with gravity}
\label{sec:gravity}
Gravity is introduced within the coset construction, following ref.~\cite{Delacretaz:2014oxa}, by  gauging the Poincar\'e group. Notice that this is a partial gauging of the full Galilean symmetry and as such it represents a (small) explicit breaking of it:\footnote{Of course there is nothing wrong in a partial gauging of the coset. A notable example is electromagnetism in the QCD chiral Lagrangian.} indeed it is well known that Galilean invariance cannot be defined in the presence of gravity. 
The coset representatives are now written as 
\begin{equation}
\begin{split}
g(y^a(\xi^\mu),\pi,\Omega^a)& =e^{y^a P_a}e^{\pi C}e^{\Omega^a B_a} \,, \\
g_{\alpha}(x^a(\xi^\mu),q,\hat{\Omega}^a) & =e^{x^a P_a}e^{q C}e^{\hat{\Omega}^a B_{(\alpha)a}} \,.
\end{split}
\end{equation}
Notice in particular that the world-line coordinates $\xi^\mu$ can always be taken to be the same, since the theory is invariant under diffeomorphisms.
The coupling with gravity corresponds to 
\be
g^{-1} {\rm d}g \rightarrow g^{-1} {\cal D} g = g^{-1} ({\rm d}+\tilde e^a P_a + \frac 12 \omega^{ab} M_{ab} ) g\,,
\ee
with $\tilde e^a = e^a - {\rm d} x^a- \omega^{ab} x_b$,  $e^a$ is the physical vierbein  defining the metric $g_{\mu \nu} = e_\mu^a e_\nu^b \eta_{ab}$ (see ref.~\cite{Delacretaz:2014oxa} for details) and $\omega^{ab}$ is the spin connection. From now on we will often omit the space-time indices, adopting a differential form notation.
At this stage, $\omega^{ab}$ is still an independent field. It is straightforward to check that this procedure takes a Galileon operator and couples it to gravity promoting standard to covariant derivatives.
The same procedure can be done in both representations and to relate them we demand the identification of the Cartan forms
\be
g^{-1} {\cal D} g = g_\alpha^{-1}  {\cal D} g_\alpha \;.
\ee
This gives the following relations:
\bea
e^a(\xi) & = & \hat e^a(\xi) + \alpha\,  \hat{D} \hat{\Omega}^a(\xi) \,,
\label{vierbeinRel} \\
\omega^{ab}(\xi)& = & \hat{\omega}^{ab}(\xi)\,,
\label{omegaRel} \\
\Omega^a(\xi) &=  & \hat{\Omega}^a(\xi) \,,
 \label{OmegaRel} 
\eea
where $\hat D \hat \Omega_a = {\rm d}\hat \Omega_a + \hat \omega_a^{\; b} \hat \Omega_b$.
As expected, these relations do not depend on the unphysical functions $x^a(\xi)$ and $y^a(\xi)$ introduced above.
The inverse Higgs constraint gives
\begin{equation}\begin{split}
{\rm d}\pi & = - e_a(\xi) \Omega^a(\xi) \,, \\
{\rm d}q & = - \hat e_a(\xi) \hat \Omega^a(\xi)  \,.
\end{split}
\end{equation}
%
By using eq.~(\ref{OmegaRel}) we have
\begin{equation}
e^{\mu a}(\xi) \frac{\partial \pi(\xi)}{\partial \xi^\mu} =\hat{e}^{\mu a}(\xi)\frac{\partial q(\xi)}{\partial \xi^\mu} \,,
\label{edpiedq}
\end{equation}
where $e^\mu$ and $\hat e^\mu$ are the inverse vierbein. In components, $e^\mu_a e_\mu^b =\hat e^\mu_a \hat e_\mu^b= \delta_a^b$.

Before discussing the gravitational sector, it is useful to see how the relations eqs.~(\ref{vierbeinRel})-(\ref{omegaRel}) reduce to eqs.~(\ref{eq.coordtransf})-(\ref{eq.piq}) when the spacetime is flat. 
The points of view taken here and in section 2 are clearly different. In section 2 we have kept the flat metric invariant, while changing the coordinates, while here we have changed the metric, as dictated by eq.~(\ref{vierbeinRel}),
and kept fixed the coordinates. Thus necessarily the comparison requires us to perform a diffeomorphism to bring the transformed metric back to its flat space form given by $\eta_{\mu\nu}$.\footnote{As we will see in the next section,
if space-time is not flat, the transformation of the metric is {\it not} a diffeomorphism.}
Let us see how this works in detail. 
If we start from Minkowski space, we get in the $\alpha$-frame the metric
\be
\hat g_{\mu\nu}(\xi) = \hat e_\mu^a(\xi) \hat e_\nu^b(\xi) \eta_{ab} = \eta_{\mu\nu} + 2 \alpha \partial_\mu\partial_\nu \pi(\xi) +\alpha^2 \partial_\mu\partial_\rho \pi(\xi) \partial_\nu\partial^\rho \pi(\xi) \,.
\label{ghat}
\ee
As expected, this is still a flat metric. By performing the coordinate transformation
\be
\xi^\mu = \xi^{\mu\prime} -\alpha \eta^{\mu\nu} \frac{\partial q(\xi^\prime)}{\partial \xi^{\nu \prime}}\,,
\label{xixiprime}
\ee
the metric (\ref{ghat}) indeed turns back into the Minkowski one: $\hat g_{\mu\nu}^\prime(\xi^\prime) = \eta_{\mu\nu}$.
This change of coordinates, eq.~(\ref{xixiprime}),
corresponds to eq.~(\ref{eq.coordtransf}), with the identification $\xi^\mu=y^\mu$, $\xi^{\prime \mu}=x^\mu$. Also eq.~(\ref{edpiedq}) equals eq.~(\ref{eq.dqdpi}).\footnote{Eq.~\eqref{eq.piq} is not independent, but just a consequence of eqs.~\eqref{eq.coordtransf} and \eqref{eq.dqdpi}.}

\subsection{Torsion}
\label{sec:torsion}

Let us analyse more closely the relations (\ref{vierbeinRel})-(\ref{omegaRel}), which look deceivingly simple. The spin connection $\omega$ is an auxiliary field whose form is obtained from its non-dynamical equations of motion \cite{Delacretaz:2014oxa}.
This corresponds to the so called first-order formulation of gravity, often also employed in the context of supergravity theories.
Finding the explicit solution for the equations of motion of $\omega$ is generally difficult, unless the theory is simple enough.
From now on, we will focus on the simplest Galilean theory coupled to gravity: the free theory of $\pi$ minimally coupled to ordinary gravity.
In the original frame, the action reads
\be
S=\frac{1}{16\pi G}  \int {\rm d}^4 x \, e R - \frac 1{2}\int {\rm d}^4 x \, e\,  e^\mu_a e^{\nu a} \partial_\mu \pi \partial_\nu \pi \,,
\label{Suntwisted}
\ee
where $e = {\rm det}(e_\mu^a)$ and
\be
R = e^\mu_a e^\nu_b R_{\mu\nu}^{ab}(\omega)\,, \ \ \ \ \ R^{ab} = {\rm d} \omega^{ab} + \omega^a_{\;\;c} \wedge \omega^{cb}\,.
\ee
In this case it is easy to find the solution for $\omega$, which coincides with the usual torsion-free condition
\be
T^a = {\rm d} e^a + \omega^a_b\wedge e^b = 0 \,.
\label{torsion1}
\ee
On the other hand, finding $\hat \omega$ from the transformed action is difficult.
It would require us to map the entire action and then extremize it  with respect to $\hat\omega$ and solve for it.
A much shorter and elegant method to get the on-shell value of $\hat\omega$ 
is obtained by mapping to the $\alpha$-frame the torsion-free condition (\ref{torsion1}).
Plugging eq.~(\ref{vierbeinRel}) in eq.~(\ref{torsion1}) gives
\bea
T^a & = & {\rm d}\left[\hat e^a+\alpha \left( {\rm d}\hat\Omega^a + \hat\omega^a_{~b} \hat{\Omega}^b \right) \right] +
{\hat\omega^a}_b \wedge \left[ \hat{e}^b+\alpha \left( {\rm d}\hat{\Omega}^b + \hat\omega^b_{~c} \hat{\Omega}^c \right) \right] \nn \\
& = & \hat{T}^a + \alpha \left(  {\rm d}\hat\omega^a_{~c}+\hat\omega^a_{~b}\wedge \hat\omega^b_{~c}\right)\hat\Omega^c =
\hat{T}^a + \alpha \hat{R}^a_{~b}(\hat \omega) \hat\Omega^b \,.
\eea
Imposing $T^a = 0$ in the original theory  gives 
\begin{equation}\label{torsion}
\hat{T}^a = - \alpha \hat{R}^a_{~b}(\hat \omega)\hat \Omega^b \,,
\end{equation}
namely we end up with a nonzero torsion in the new theory. The spin-connection $\hat \omega$ has then a contorsion term $K_\mu^{ab}$ (see, i.e., chapter 7 of ref.~\cite{FP} for the basics of differential geometry including torsion).
In components,
\begin{equation}\label{contorsion1}
\hat \omega_\mu^{ab}=\bar{\omega}_\mu^{ab}(\hat e)+ \hat K_\mu^{ab} \,,
\end{equation}
where $\bar{\omega}_\mu^{ab}(\hat e)$ is the usual torsion-free part of the connection and
\begin{equation}\label{contorsion2}
\hat K_{\mu~\rho}^{~\nu}\equiv \hat K_\mu^{ab} \hat e^{\nu}_b  \hat e_{\rho a}  = 
-\frac{1}{2}\left( 
\hat T_{\mu~\rho}^{~\nu}-\hat T_{~\rho\mu}^{\nu}+\hat T_{\rho\mu}^{~~\nu} \right) \,,
\end{equation}
in terms of the torsion tensor 
\begin{equation}
\hat T_{\mu\nu}^{~~\rho}\equiv \hat T_{\mu\nu}^a \,\hat e^\rho_a\,.
\end{equation}
Using eq.~(\ref{contorsion1}) and the definition of curvature two-form, one has
\be
 \hat{R}^{a}_{\;\; b}(\hat \omega)=\hat{R}^{a}_{\;\; b}(\bar \omega)+\bar D \hat K^{a}_{\;\; b} + \hat K^{a}_{\;\; c}\wedge \hat K^{c}_{\;\; b}\,,
 \label{Rtwoform}
\ee
where $\bar D \hat K^a_{\;\; b} \equiv {\rm d} \hat K^a_{\;\; b} + 2 \;\bar \omega^a_{\;\; c} \wedge \hat K^c_{\;\; b}$. 

Finding the explicit form of $\hat T$ in terms of $\hat e$ using eq.~(\ref{torsion}) is non-trivial, since the torsion appears also on the right-hand side.
Of course when we start with zero curvature in the original frame, both curvature and torsion are zero in the $\alpha$-frame.
In general one can recursively get a perturbative expression of $\hat T$ in the regime $\alpha\partial^2 q \ll 1$, $\alpha\partial q \partial R \ll R$.
The resulting expression is an infinite series of terms involving $\hat R^a_{\;\; b}(\bar \omega)\hat \Omega^b$ and its derivatives.
We will see that a subset of such terms is crucial in discussing the graviton propagation in a galileon background.

\subsection{The Dual Action}
\label{sec:dualaction}

In this section we want to see how the action (\ref{Suntwisted}) is transformed under the mapping (\ref{vierbeinRel})-(\ref{omegaRel}).
Let us first consider the second term in eq.~(\ref{Suntwisted}). Thanks to eq.~(\ref{edpiedq}), the kinetic term of $\pi$ is simply mapped to that of $q$.
The non-trivial action arises from the mapping of $e$. When expressed in terms of $\hat e$,  we get the 4 Galileon terms found in refs.~\cite{deRham:2013hsa,Creminelli:2014zxa}, dressed with gravitational interactions.
The important point to notice is that the connection appearing in the covariant derivative is torsionful, namely one has
\be
 \nabla_\mu \nabla_\nu q = \partial_\mu \partial_\nu q - \Gamma_{\mu~\nu}^{~\rho} \partial_\rho q \,,
\ee
where
\begin{equation}
\Gamma_{\mu~\nu}^{~\rho}=\bar\Gamma_{\mu~\nu}^{~\rho}+K_{\mu~\nu}^{~\rho}\,,
\end{equation}
$\bar{\Gamma}$ is the Levi-Civita connection and $K$ is the contorsion tensor defined in terms of the torsion $T$ in eq.~(\ref{contorsion2}).
So, while in flat space with no gravity the action of a free scalar is mapped to a sum of a finite number of terms, corresponding to the 4 galileon 
interactions, in the presence of gravity an infinite number of higher-derivative terms are induced in the action. These come from the recursive solution for the torsion discussed at the end of the previous section.

Let us now consider how the map acts on the Einstein-Hilbert term in eq.~(\ref{Suntwisted}), that we rewrite as 
\begin{equation}
S_{EH}=\frac{1}{32\pi G}  \int \epsilon_{abcd} \, e^a\wedge e^b \wedge R^{cd}(\omega) \,.
\end{equation}
Using eqs.~(\ref{vierbeinRel})-(\ref{omegaRel}) we immediately get the action in the $\alpha$-frame:
\begin{equation}\label{gravity}
\hat S_{EH}=\frac{1}{32\pi G}   \int \epsilon_{abcd} \left(\hat e^a\wedge \hat e^b + 2\alpha \hat D\hat \Omega^a \wedge \hat e^b + \alpha^2 \hat D\hat \Omega^a \wedge \hat D\hat \Omega^b \right)  \wedge \hat R^{cd}(\hat \omega)\,,
\end{equation}
where $\hat D$ is the complete $\hat \omega$-covariant derivative, which includes the contorsion term.
In order to simplify the notation, from now on we will omit the hatted indices.
It is useful to keep in mind the following properties to manipulate the action (\ref{gravity}):
\be
{\rm d}\beta= D\beta\,, \ \ \ \  \int {\rm d}\beta = 0 \,, \ \ \ \ 
DD\Omega^i  =R^i_{~j}\Omega^j \,, \ \ \ \ \ 
DR^{ij}  =0 \,, \ \ \ \ \ 
De^a  =T^a \,,
\end{equation}
where $\beta$ is any $SO(3,1)$ singlet 3-form.
Integrating by parts the second and third terms, we get
\be
\hat S_{EH}=\frac{1}{32\pi G} \int \epsilon_{abcd} \left( e^a\wedge e^b - 2\alpha \Omega^a T^b - \alpha^2 \Omega^a R^b_{~k}\Omega^k \right)  \wedge R^{cd}(\omega)\,.
\label{ShatEH}
\ee
Since $T$ and $K$ are at least linear in $\alpha$, see eq.~(\ref{torsion}),  it is useful to define  $K_\mu^{ab}= \alpha \mathcal{K}_\mu^{ab}$. 
Using eqs.~(\ref{torsion}) and (\ref{Rtwoform}), we have
\begin{equation}
\hat S_{EH}=\frac{1}{32\pi G}\int \epsilon_{abcd} \left( e^a\wedge e^b + \alpha^2 \Omega^a R^b_{~k}(\omega)\Omega^k \right)  \wedge 
\left( R^{cd}(\bar{\omega})  + \alpha \bar{D}\mathcal{K}^{cd} + \alpha^2 \mathcal{K}^c_{~i}\wedge \mathcal{K}^{id}  \right) \,.
\end{equation}
The only term that appears at linear order in $\alpha$ is a total derivative since $\bar{D}e^a=0$. We are then left with
\be
\hat S_{EH}  = \frac{1}{32\pi G} \epsilon_{abcd} \int e^a\wedge e^b \wedge R^{cd}(\bar{\omega})+
\alpha^2
\Big( e^a \wedge e^b \wedge \mathcal{K}^c_{~m} \wedge \mathcal{K}^{md} + \Omega^a R^b_{~k}(\omega)\Omega^k \wedge R^{cd}(\omega) \Big) \,.
\label{ShatEHfinal}
\ee
The action (\ref{ShatEHfinal}) is a sum of a finite number of terms, but each term (aside from the first) contains an infinite number of terms induced by the torsion (\ref{torsion}). Hence Einstein gravity in one frame is mapped to a complicated theory featuring an infinite number of higher-derivative terms.

\section{Superluminality and conclusions}
\label{sec:conclusions}
Our original motivation was to understand whether the Galileon duality, and in general any reparametrization of a space-time coset, leaves the physics invariant. The coupling of the theory with gravity shows that if we start with a theory coupled to gravity in a ``standard'' way, the reparametrization leads to a complicated action for gravity with an infinite series of higher dimension operators suppressed by a scale much lower than $M_{\rm Pl}$. When all these operators become relevant the dual reformulation of the theory is effectively non-local. We can study when this happens on a background for $q$; for simplicity let us assume that the background is slowly varying: $\alpha \partial^2 \bar q \ll 1$. The recursion given by eq.~\eqref{torsion} contains terms of the form
\be
\alpha^n \partial^{n-1} R (\partial \bar q)^{n} \,,
\label{TermsSuperluminal}
\ee
where we neglected terms with more than one derivative acting on $\bar q$.
If one now considers a process with typical frequency $\omega$, for example the propagation of a gravitational wave, the series of terms above cannot be truncated when 
\be
\alpha \omega \partial \bar q \gtrsim 1 \;.
\ee
In this regime all operators contribute in the same way and the theory in the new parametrization is effectively non-local. It is important to stress that the condition above coincides with the condition of measurable superluminality of the $q$ fluctuations around the $\bar q$ background induced, even in the absence of gravity, by the Galileon terms.\footnote{In the way the reparametrization of the coset is done in this paper, the coordinates $\xi^\mu$ are not touched, so that in the two frames the worldline $\xi^\mu(\lambda)$ describing how perturbations propagate is exactly the same. However, the metric changes so that the same worldline will be seen as superluminal or subluminal in the two frames. In the flat space treatment of ref.~\cite{Creminelli:2014zxa} the metric remains $\eta_{\mu\nu}$ but the coordinates change modifying whether the propagation is sub- or super-luminal. As we discussed above the two approaches are simply related by a diffeomorphism in flat space.} When superluminality is measurable the mapping between the theories is clearly non-local since an infinite set of higher-dimension operators becomes relevant. 

Summarizing, {\it Galileon theories with a standard coupling with gravity are not mapped to each other, but to theories in which gravity contains a series of higher dimension operators suppressed by a low scale.} In a theory with a standard coupling with gravity, terms of the form (\ref{TermsSuperluminal}) can be generated at the quantum level, but at parametrically higher scales, governed by inverse powers of $M_{\rm Pl}$. 
In particular they should vanish in the limit $M_{\rm Pl}\rightarrow \infty$. In this sense the theory featuring the operators (\ref{TermsSuperluminal}) is not a standard theory of gravity.\footnote{We thank Enrico Trincherini for 
discussions on this point.} The equivalence among parametrizations is thus broken once one specifies in which one the coupling with gravity is standard.

We think our results apply straightforwardly to other cosets. For example in the case of the breaking of the conformal group SO(4,2) to the Poincar\'e group, there are (at least) two interesting parametrizations. In this case the mapping between the two is not an automorphism, but relates two genuinely different ways of non-linearly realizing the conformal group \cite{Bellucci:2002ji,Creminelli:2013fxa}. In each representation there is a set of five conformal Galilean operators which are mapped one into the other by the reparametrization of the coset \cite{Creminelli:2013fxa}. Also in this case we expect that if one chooses a standard coupling with gravity in one of the two representations, this will correspond to a weird coupling in the other. Both sets of Galilean operators were used as candidates for cosmological scenarios which are alternative to inflation \cite{Creminelli:2010ba,Hinterbichler:2012yn}. In each case Galileons were coupled to gravity in a standard way. Therefore we now understand that the two scenarios are not equivalent, since they have a different coupling with gravity. This conclusion agrees with the fact that superluminality constraints are very different in the two cases.

It would be interesting to understand whether our results are also relevant in cases where the unbroken group does not contain Poincar\'e, like in the case of ordinary fluids and solids.

\subsection*{Acknowledgements}
We are indebted to Alberto Nicolis, Riccardo Penco and Andrew J. Tolley for useful discussions and especially to Enrico Trincherini for early collaboration on this project.
The work of M.S. was supported by the ERC Advanced Grant no. 267985 DaMESyFla.

\end{document}